\documentclass{XrU2005}
\usepackage{mathptm}
\usepackage{graphicx}

\title{Gaseous abundances in M82}
\author{P. Ranalli}
\affil{RIKEN, Cosmic Ray Laboratory, 2-1 Hirosawa, Wakoshi, Saitama,
  351-0106 Japan; JSPS fellow}
\author{L. Origlia$^2$, A. Comastri}
\affil{INAF, Osservatorio Astronomico di Bologna, via Ranzani 1, 40127
  Bologna, Italy}
\author{R. Maiolino}
\affil{INAF, Osservatorio Astrofisico di Arcetri, L.go E. Fermi 5,
  50125 Firenze, Italy}
\author{K. Makishima$^{1,}$}
\affil{Department of Physics, The University of Tokyo, 7-3-1 Hongo,
  Bunkyo-ku, Tokyo 113-0033}

\newcommand{\xmm}{XMM-{\em Newton}}
\newcommand{\chandra}{{\em Chandra}}

\newcommand{\ael}{$\alpha$-element}
\newcommand{\lesssim}{<\sim}

\begin{document}


\maketitle

\section{Introduction}
The signature of the star formation (SF) history of a galaxy is
imprinted in the abundance patterns of its stars and gas.  Determining
the abundance of key elements released in the interstellar medium
(ISM) by stars with different mass progenitors and hence on different
time scales, will thus have a strong astrophysical impact in drawing
the global picture of galaxy formation and evolution (McWilliam, 1997,
ARA\&A, 35, 503).  It also offers the unique chance of directly
witnessing the enrichment of the ISM (Maeder \& Conti, 1994, ARA\&A,
32, 227).  Metals locked into stars give a picture of the enrichment
just prior to the last burst of SF, while the hot gas heated by SNe~II
explosions and emitting in the X-rays should trace the enrichment by
the new generation of stars.

We have started a project to measure the metallicity enhancement in a
sample of starburst galaxies, for which we obtained high resolution
infrared (J and H band) spectra with the 3.6 m Italian Telescopio
Nazionale Galileo (TNG) and with the ESO VLT, and both proprietary and
archival data from the \xmm\ and \chandra\ missions.  Our sample
comprises M82, NGC253, NGC4449, NGC3256 and the {\em Antennae},
sampling two orders of magnitude in star formation, as it ranges from
the 0.3 $M_\odot$/yr of NGC4449 to the $\sim 30 M_\odot$/yr of the
Antennae and NGC3256.

Preliminary results for M82 were achieved with the available \xmm\ 
archival data, and hinted for a confirmation of the expected scenario
in which the gaseous component has a higher content of
$\alpha$-elements than the stellar one, and a similar content of Fe
(Origlia et al., 2004, ApJ 606, 862).  However, some new issues were
posed, since we found a very low abundance of O and Ne with respect to
other $\alpha$-elements (e.g., O/Mg $\sim 0.2$, Ne/Mg $\sim 0.3$) in
the hot gas present in the central ($\lesssim 1$ kpc) regions of M82,
which could not be satisfactorily explained. Thus we were granted a
deeper observation of M82, whose preliminary results are presented in
the following.

\setcounter{footnote}{-1}  

\begin{figure*}[t]
\centering
\includegraphics[width=.85\columnwidth]{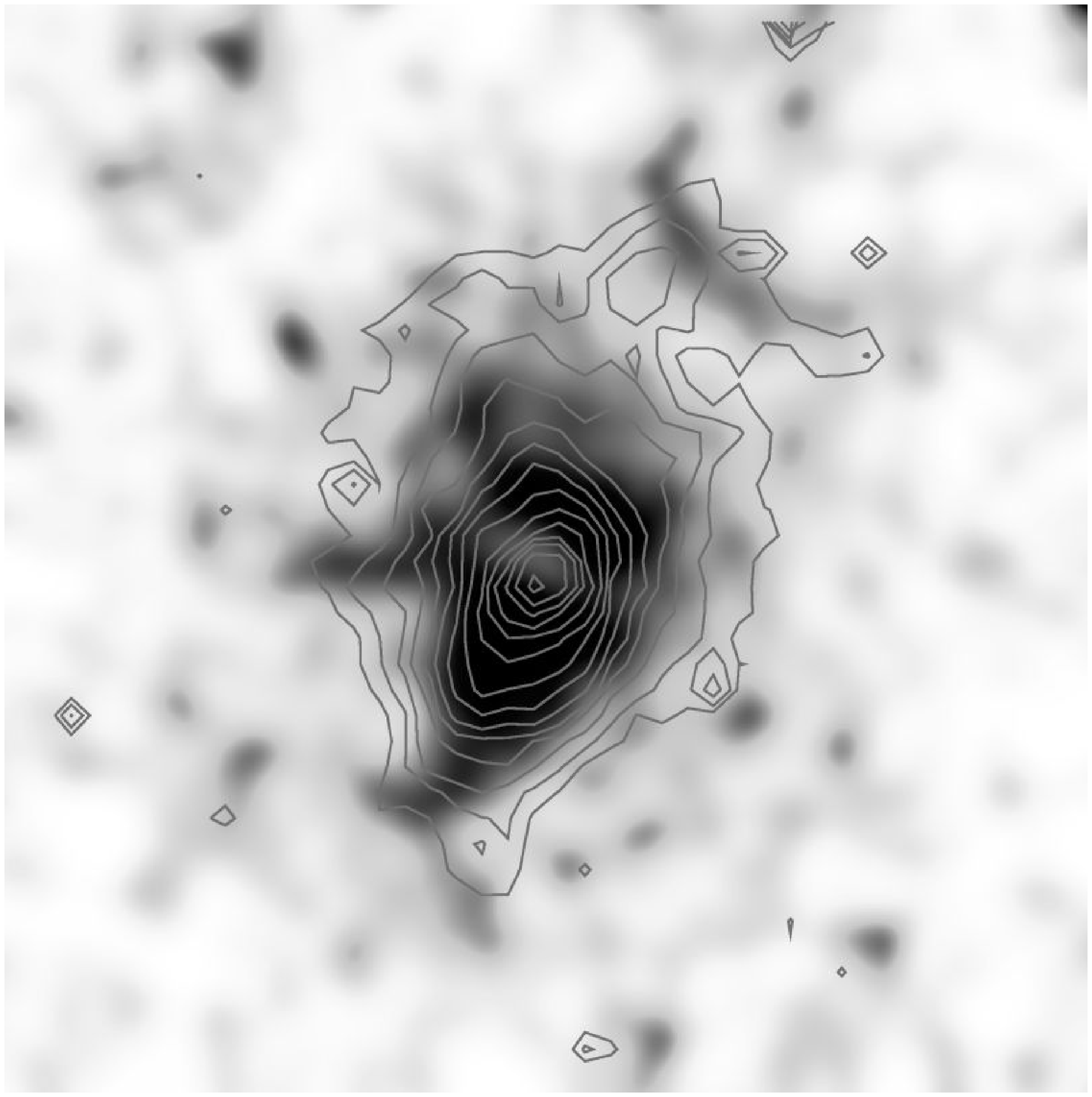}
\hspace{1cm}
\includegraphics[width=\columnwidth]{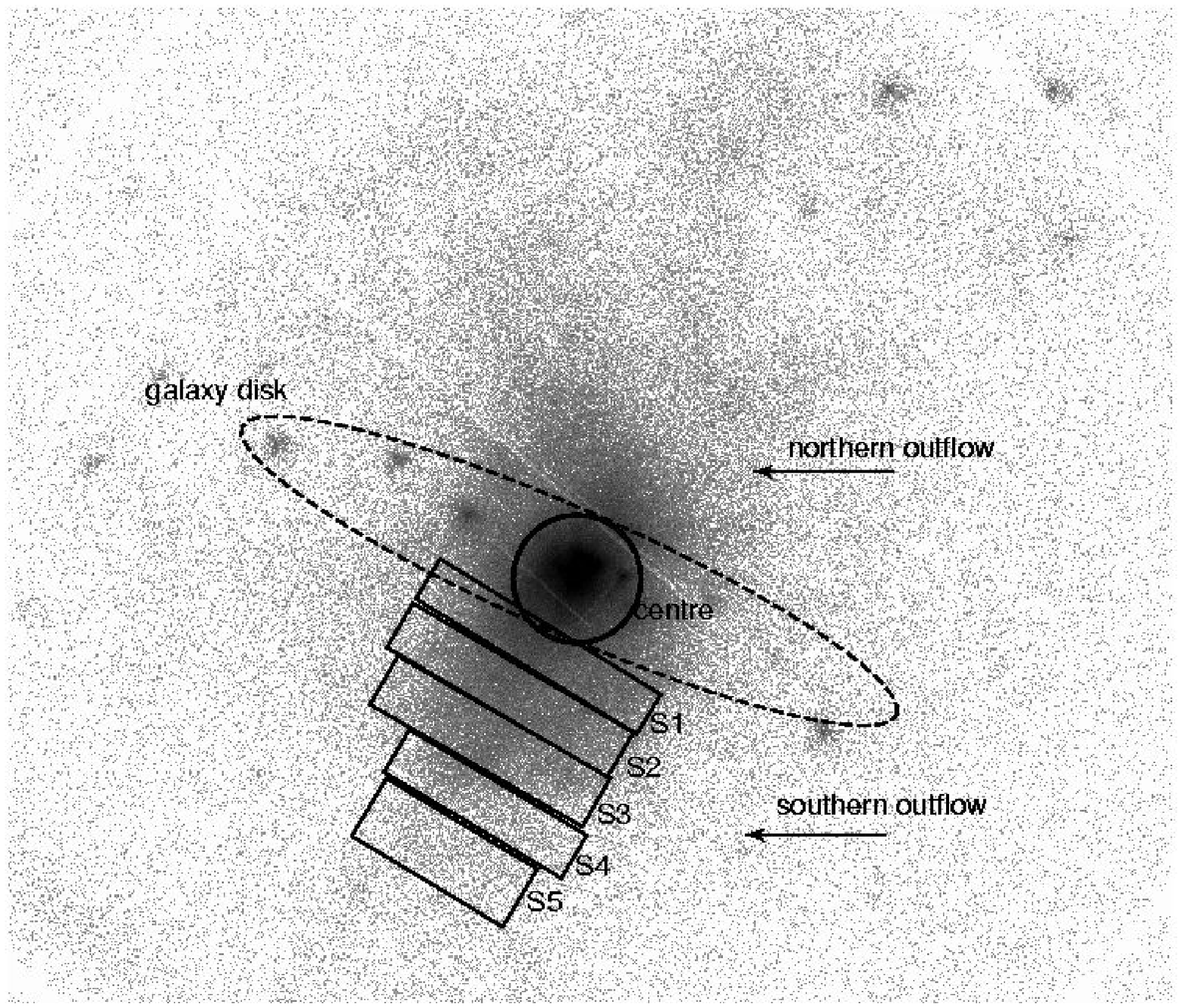}
\caption{{\bf Left:} Ne X abundance map.
  Gray scale: line/continuum flux ratio, tracking Ne abundance; darker
  means more abundant.
  Contours: 0.5--2.0 keV emission.  {\bf Right:} Image of 0.5--10 keV
  emission with a sketch of the regions used in the spectroscopical
  analysis.
\label{fig:m82neon}}
\end{figure*}
\begin{table*}[t]
  \centering
  \label{tab:abbondanze1}
  \caption{Chemical abundances in different regions with 90\%
    errors. The abundances are given in solar units,
    i.e. Fe/Fe$_\odot$, following the Grevesse \& Sauval (1998,
    Sp.Sci.Rev.\ 85, 161) scale. Region S4 is not shown since it falls mainly
    on CCD gaps. The ``stars'' region shows the results from infrared
    spectroscopy, while the ``centre'' region is referred to the
    PN+RGS data analysis in Origlia et al.\ (2004).
    The average height above the galactic plane
    is reported, assuming a distance of 2.94 Mpc (de Vaucouleurs et
    al., ``Third reference catalogue of bright galaxies'', 1991,
    assuming $H_0=70$).}
  \begin{tabular}{cccccccc}
  \hline
  region &height (kpc) &Fe &O &Ne &Mg &Si &S \\
  \hline
  centre &       &$0.43_{-0.08}^{+0.12}$  &$0.26_{-0.09}^{+0.15}$ 
                 &$0.45_{-0.12}^{+0.17}$  &$1.36_{-0.26}^{+0.32}$
                 &$1.49_{-0.26}^{+0.32}$  &$1.42_{-0.40}^{+0.48}$
  \smallskip\\
  S1     &0.6    &$0.48\pm 0.01$  &$0.57\pm 0.04$  &$0.85\pm 0.05$
                 &$1.33\pm 0.04$  &$1.12\pm 0.05$  &$0.75\pm 0.10$\\
         
  S2     &1.0    &$0.57\pm 0.02$  &$0.64\pm 0.04$  &$1.26\pm 0.07$
                 &$1.73\pm 0.07$  &$1.33\pm 0.09$  &$0.75\pm 0.17$\\

  S3     &1.5    &$0.88\pm 0.04$  &$1.04\pm 0.07$  &$2.21\pm 0.14$
                 &$3.06\pm 0.15$  &$2.06\pm 0.17$  &$1.29\pm 0.33$\\


  S5     &2.5    &$0.78\pm 0.08$  &$1.71\pm 0.16$  &$3.31\pm 0.35$
                 &$4.09\pm 0.48$  &$2.65\pm 0.74$  &$0.53_{-0.53}^{+1.93}$
  \smallskip\\         
  stars  &       &$0.5\pm 0.2$ &$1.0_{-0.3}^{+0.5}$  &---
                 &$1.1_{-0.3}^{+0.4}$  &$1.1_{-0.5}^{+1.0}$ &---\\

\end{tabular}
\end{table*}

\section{Mapping the chemical elements through narrow-band imaging}
\label{sec:morpho}
The MOS and PN cameras onboard \xmm\ have a moderate energy resolution
(FWHM $ \sim 70-80$ eV at 1 keV), which allows the use of narrow-band
imaging to infer the distribution of chemical elements throughout the
galaxy. Among the most prominent spectral lines lying around $\sim 1$
keV, we consider here the Ne X line at 10~\AA\ because this element,
together with O, posed the main problems in our previous work.
Unfortuately, it is not possible to consider lines from O, since the
MOS resolution is rapidly degraded at energies $\lesssim 0.6$ keV, and
the enlargement of the band needed to take account of the instrumental
effects would make an image in the O band seriously contaminated
by continuum emission.

In order to extract information about aboundances, the line-band
images should be normalized by accounting for the continuum
emission. We perform this correction by  dividing each line-band image
by a smoothed 0.5--2.0 keV band image. Results from spacially-resolved
spectroscopy confirmed that once this correction is applied, the
line-band images do trace the chemical abundances.

Fig.~\ref{fig:m82neon} (left panel) shows the Ne abundance
distribution map with superimposed contours from 0.5-2.0 keV emission.
Ne is clearly concentrated in two separated regions, north and south
of the galaxy center.

\section{Spatially resolved spectroscopy}
\label{sec:spectro}
We present here the preliminary analysis conduced on the southern
outflow, making use of EPIC data.  We divided the southern outflow in
five regions, in order to study the different properties of the hot
gas as it flows and/or is heated from the central starburst towards
the intergalactic space. The regions are numbered from S1 to S5 with
increasing height above the galactic plane, and are shown in
Fig.~\ref{fig:m82neon} (right panel). The spectra were extracted from
the MOS (0.5--8.0 keV) and PN (1.0--9.0 keV) data, and fitted with a
multi-temperature ``mekal'' thermal plasma model whose differential
emission measure (DEM) is described by a 6$^{\rm th}$ order polynomial
in the 0.1--10 keV energy range. Background spectra were extracted
from the blank-sky data files (Read \& Ponman, 2003, A\&A, 409, 395)
after normalization to the background levels observed in the M82 data.

The best-fit chemical abundances are shown in
Table~\ref{tab:abbondanze1} along with results form our previos paper
(Origlia et al., 2004, {\em op.\ cit.}) relative to X-ray (EPIC and
RGS) and infrared data for the central regions, marked in the Table as
``centre'' and ``stars'', respectively. No significant changes are
found in the temperature of the plasma. The only region whose spectra
require absorption in excess of the Galactic value is S1. Our previous
finding, that the inner region of M82 is somewhat devoid of the
lighter \ael, is thus confirmed. Moreover, it is found that these
elements are rather to be concentrated in the outflow.  On the other
hand, the heavier elements (Mg, Si, S) while following a similar
spatial pattern seem to be more evenly distributed.

\section{Discussion}
\label{sec:discu}
In M82 both the hot gas and the stellar phases trace a very similar Fe
abundance. 
Indeed, since Fe is mainly produced by type Ia supernovae (SN), it is
expected to be released in the ISM only after $\sim 1$ Gyr from the
local onset of star formation. At variance, \ael\ (O, Ne, Mg, Si, S)
are predominantly released by type II SN with massive progenitors on
much shorter time scales. The overall \ael/Fe enhancement in the
innermost region of M82 is consistent with a standard chemical
evolution scenario only for heavier elements (Mg, Si, S).

However, the lighter elements (O, Ne) show a different distribution,
in which the inner regions of the galaxy appears somewhat devoid of
these metals, while in outer parts of the outflow they are found to be
enhanced as the heavier \ael. 
When the entire sample of galaxies will be analyzed, we will better
understand whether the bipolar distribution of O and Ne is peculiar of
M82 or is a common feature in starburst galaxies.

\end{document}